\documentclass[aps,prb,twocolumn,showpacs]{revtex4}

\usepackage{graphicx}
\usepackage{amsmath}
\usepackage{amssymb}

\begin{document}

\title{ Quantum phase transition in the one-dimensional compass model }

\author{      Wojciech Brzezicki }
\affiliation{ Marian Smoluchowski Institute of Physics, Jagellonian 
              University, Reymonta 4, PL-30059 Krak\'ow, Poland  }

\author{      Jacek Dziarmaga }
\affiliation{ Marian Smoluchowski Institute of Physics, Jagellonian 
              University, Reymonta 4, PL-30059 Krak\'ow, Poland  }
\affiliation{ Centre for Complex Systems, Jagellonian University,
              Reymonta 4, PL-30059 Krak\'ow, Poland  }

\author{      Andrzej M. Ole\'s }
\affiliation{ Marian Smoluchowski Institute of Physics, Jagellonian 
              University, Reymonta 4, PL-30059 Krak\'ow, Poland  }
\affiliation{ Max-Planck-Institut f\"ur Festk\"orperforschung, 
	      Heisenbergstrasse 1, D-70569 Stuttgart, Germany }

\date{ \today }

\begin{abstract}
We introduce a one-dimensional model which interpolates between the 
Ising model and the quantum compass model with frustrated pseudospin 
interactions $\sigma_i^z\sigma_{i+1}^z$ and $\sigma_i^x\sigma_{i+1}^x$, 
alternating between even/odd bonds, and present its exact solution by 
mapping to quantum Ising models. 
We show that the nearest neighbor pseudospin correlations 
change discontinuosly and indicate divergent correlation length 
at the first order quantum phase transition. At this transition 
one finds the disordered ground state of the compass model with 
high degeneracy $2\times 2^{N/2}$ in the limit of $N\to\infty$. 
\end{abstract}

\pacs{75.10.Jm, 03.67.Lx, 05.70.Fh, 73.43.Nq}

\maketitle

\section{Introduction}

Recent interest in quantum models of magnetism with exotic interactions 
is motivated by rather complex superexchange models which arise in Mott 
insulators with orbital degrees of freedom.\cite{Tok00} The 
degeneracy of $3d$ orbitals is only partly lifted in an octahedral 
environment in transition metal oxides, and the remaining orbital 
degrees of freedom are frequently described as $T=1/2$ pseudospins. 
They occur on equal footing with spins in spin-orbital superexchange 
models and their dynamics may lead to enhanced quantum fluctuations near 
quantum phase transitions,\cite{Fei97} and to entangled spin-orbital
ground states.\cite{Ole06} The properties of such models are still
poorly understood, so it is of great interest to investigate first the 
consequences of frustrated interactions in the orbital sector alone.

The orbital interactions are intrinsically frustrated ---
they have much lower symmetry than the usual SU(2) symmetry of spin 
interactions, and their form depends on the orientation of the bond in 
real space,\cite{vdB04} so they may lead to orbital liquid in three 
dimensions.\cite{Kha00} Although such interactions are in reality more 
complicated,\cite{Fei97,Ole06,vdB04,Kha00} a generic and simplest 
model of this type is the so-called compass model introduced long ago, 
\cite{Kho82} when the coupling along a given bond is Ising-like, but 
different spin components are active along different bond directions, 
for instance 
$J_x\sigma_i^x\sigma_j^x$ and 
$J_z\sigma_i^z\sigma_j^z$ 
along $a$ and $b$ axis in the two-dimensional (2D) compass model. 
The compass model is challenging already for the classical interactions.
\cite{Nus04} Although the compass model originates from the orbital 
superexchange, is is also dual to recently studied models of $p+ip$ 
superconducting arrays.\cite{Nus05} It was also proposed as a realistic 
model to generate protected cubits,\cite{Dou05} so it could play a role 
in the quantum information theory. 

So far, the nature of the ground state and pseudospin correlations in 
the compass model were studied only by numerical methods.
It has been argued that the eigenstates of the 2D compass model are 
twofold degenerate.\cite{Dou05} In contrast, a numerical study of the 
2D compass model suggests that the ground state is highly degenerate 
and disordered.\cite{Mil05} In fact, the numerical evidence suggests 
a first order quantum phase transition at $J_x=J_z$,\cite{Kho03} 
with diverging spin fluctuations and a discontinuous change in the 
correlation functions.   

The purpose of this paper is to show by an exact solution a mechanism 
of a first order quantum phase transition in quantum magnetism. Such 
a transition from quasiclassical states with short range order to a 
disordered state occurs in the one-dimensional (1D) XX--ZZ model, 
with antiferromagnetic interactions \cite{noteaf} between $\sigma_i^x$ 
and $\sigma_i^z$ pseudospin components, alternating on even/odd bonds. 
The model is solved analytically by mapping its different subspaces 
to the exactly solvable quantum Ising model (QIM),\cite{Lie61} which 
plays a prominent role in understanding the paradigm of a quantum phase 
transition, including recent rigorous insights into the transition 
through its quantum critical point at a finite rate.\cite{Dzi05} 
Note that the model introduced below is generic and by no means limited 
to the orbital physics. For instance, the QIM is realized by the 
charge degrees of freedom in NaV$_2$O$_5$, where it helped to explain 
the temperature dependence of optical spectra.\cite{Aic02}

The paper is organized as follows: In Sec. II we introduce the 
pseudospin (orbital) model which realizes a continuous interpolation 
between the classical Ising model and the compass model in one 
dimension. This model is next solved exactly in its subspaces which 
are separated from each other. The properties of the model and the 
mechanism of the quantum phase transition are elucidated in Sec. III. 
At the transition point we demonstrate the discontinuous change of 
correlation functions (Sec. III.A) and the vanishing pseudospin gap 
(Sec. III.B). The paper is summarized in Sec. IV.

\section{Pseudospin XX--ZZ model}

In order to understand the nature of a quantum ground state with high 
degeneracy found in the 1D compass model with alternating superexchange 
interactions between $\sigma_i^x$ and $\sigma_i^z$ pseudospin 
components, it is important to investigate how the pseudospin 
correlations develop when this state is approached. Therefore, we start 
below from the classical Ising model with interacting $\sigma_i^z$ 
pseudospins and modify gradually the interactions on every second bond 
by replacing them by the ones between $\sigma_i^x$. Although this does 
not correspond to any deformation of interacting orbitals in a crystal, 
in this way we keep a constant value of the total pseudospin coupling 
constant $J$ on each bond, and can demonstrate that the frustration of 
interactions increases and dominates the behavior of the compass model
at the quantum critical point.    

\subsection{Hamiltonian and its subspaces}

We consider the pseudospin model with interactions depending on
parameter $\alpha\in[0,2]$. When $0\leq\alpha\leq 1$
\begin{equation}
{\cal H}(\alpha)\equiv J\sum_{i=1}^{N'} 
\left[ (1-\alpha) \sigma_{2i-1}^z \sigma_{2i}^z + 
          \alpha  \sigma_{2i-1}^x \sigma_{2i}^x +  
	          \sigma_{2i}^z   \sigma_{2i+1}^z \right], 
\label{Halpha}
\end{equation}
with $N'=N/2$, but when $1\leq\alpha\leq 2$
\begin{eqnarray}
{\cal H}(\alpha)&\equiv&J
\sum_{i=1}^{N'} 
\big[ \sigma_{2i-1}^x \sigma_{2i}^x                   \nonumber \\
&+&  (2-\alpha) \sigma_{2i}^z \sigma_{2i+1}^z + 
     (\alpha-1) \sigma_{2i}^x \sigma_{2i+1}^x \big]. 
\label{Halpha12}
\end{eqnarray}
Here $\{\sigma_i^x,\sigma_i^z\}$ are Pauli matrices. We assume even 
number of pseudospins $N$ and periodic boundary conditions: 
$\sigma_{N+1}=\sigma_1$. The model reduces to the ZZ Ising model
at $\alpha=0$, 
\begin{equation}
{\cal H}(0)=J\sum_{i=1}^N \sigma_i^z \sigma_{i+1}^z\,, 
\label{H0}
\end{equation}
and to the XX Ising model at $\alpha=2$,
\begin{equation}
{\cal H}(2)=J\sum_{i=1}^N \sigma_i^x \sigma_{i+1}^x\,.
\label{H2}
\end{equation}
It is frustrated between these two incompatible types of order for 
$0<\alpha<2$, and we focus on the consequences of increasing 
frustration. Right in the middle between the two classical cases 
(\ref{H0}) and (\ref{H2}), i.e. for $\alpha=1$, it becomes the proper 
quantum XX--ZZ model (called {\it 1D compass model\/}), 
\begin{equation}
{\cal H}(1)=J\sum_{i=1}^{N'} 
\left[ \sigma_{2i-1}^x \sigma_{2i}^x +
       \sigma_{2i}^z   \sigma_{2i+1}^z \right]~,
\label{H1}
\end{equation}
favoring antiferromagnetic order of $x$ and $z$ pseudospin components 
on odd and even bonds, respectively. 

The Hamiltonians (\ref{Halpha}) and (\ref{Halpha12}) are related by 
symmetry: simultaneous exchange of 
$\sigma_i^x\leftrightarrow\sigma_i^z$,
$2i\leftrightarrow 2i-1$ $\forall$ $i$, and 
$(1-\alpha)\leftrightarrow(\alpha-1)$, 
maps the Hamiltonians into each other. 
Thanks to this symmetry, we need to solve only the model (\ref{Halpha}) 
for $\alpha\in[0,1]$, with modulated interactions for {\it odd pairs\/} 
of pseudospins $\{2i-1,2i\}$ (on {\it odd bonds\/}). 
The Hamiltonian (\ref{Halpha}) can be conveniently solved in the basis 
of eigenstates of the ZZ Ising model at $\alpha=0$. We start with one 
of the two degenerate ground states,
\begin{equation}
\left|
\uparrow\downarrow\uparrow\downarrow\uparrow\downarrow
\uparrow\downarrow\uparrow\downarrow\uparrow\downarrow
\dots\right\rangle ~,
\label{GS0}
\end{equation} 
which is modified at finite $\alpha>0$ by spin-flip operators: 
$-\tau^x_i\equiv\sigma_{2i-1}^x\sigma^x_{2i}$. Action of various 
$\tau^x$'s on the state (\ref{GS0}) generates other eigenstates:
\begin{eqnarray}
&\left|
\downarrow\uparrow\uparrow\downarrow\uparrow\downarrow
\uparrow\downarrow\uparrow\downarrow\uparrow\downarrow
\dots\right\rangle&, \nonumber\\
&\left|
\uparrow\downarrow\downarrow\uparrow\uparrow\downarrow
\uparrow\downarrow\uparrow\downarrow\uparrow\downarrow
\dots\right\rangle&, 
\;\dots~, \nonumber\\
&\left|
\downarrow\uparrow\downarrow\uparrow\uparrow\downarrow
\uparrow\downarrow\uparrow\downarrow\uparrow\downarrow
\dots\right\rangle&, 
\;\dots~, \nonumber\\
&\left|
\downarrow\uparrow\downarrow\uparrow\downarrow\uparrow
\downarrow\uparrow\downarrow\uparrow\downarrow\uparrow
\dots\right\rangle&, 
\label{GSsubspace}
\end{eqnarray}  
where the last state is the second degenerate ground state of the ZZ 
Ising model. This set of states (\ref{GS0})--(\ref{GSsubspace}) is a 
convenient basis in the subspace where all odd pairs of pseudospins 
are antiparallel, and give a constant energy contribution of 
$-(1-\alpha)J$ per bond. By construction, the Hamiltonian (\ref{Halpha}) 
does not mix this subspace with the rest of the Hilbert space. 
Therefore, in this subspace the energy due to the {\it even\/} 
$\{2i,2i+1\}$ bonds in Eq. (\ref{Halpha}) depends on the pseudospin 
orientation on the two neighboring odd bonds, and may be expressed 
by a product $\tau^z_i\tau^z_{i+1}$ of two spin operators,  
with $-\tau^z_i\equiv\sigma_{2i-1}^z\sigma_{2i}^z$. 
In this representation it is clear that the Hamiltonian (\ref{Halpha})
reduces in this subspace to the exactly solvable QIM.\cite{Lie61}

In general the Hilbert space can be divided into subspaces where
different odd pairs of spins are either parallel or antiparallel.
Each subspace can be labelled by a vector $\vec s=(s_1,...,s_{N'})$,
with $s_i=1$ ($s_i=0$) when two pseudospins of the odd bond 
$\{2i-1,2i\}$ are parallel (antiparallel). In each subpace $\vec s$ 
the terms $\propto (1-\alpha)$ for odd bonds in Eq. (\ref{Halpha}) 
give a constant energy contribution
\begin{eqnarray}
C_s(\alpha)&\equiv& (1-\alpha)J\sum_{i=1}^{N'}
\sigma_{2i-1}^z\sigma_{2i}^z                       \nonumber \\
&=&-(1-\alpha)J(N'-2s)\,, 
\label{E0s}
\end{eqnarray}
where $s=\sum_{i=1}^{N'}s_i$ is the number of parallel odd pairs of 
spins. With a convenient choice of spin operators for the antiparallel 
$\{2i-1,2i\}$ pairs of pseudospins,
\begin{eqnarray}
-\tau^x_i&=& 
\left( \left|\uparrow\downarrow\rangle
        \langle\downarrow\uparrow\right|+
       \left|\downarrow\uparrow\rangle
        \langle\uparrow\downarrow\right|\right)~,
\nonumber\\    
-\tau^z_i&=& 
(-1)^{\sum_{j=1}^{i-1} s_i}~
\left( \left|\uparrow\downarrow\rangle
           \langle\uparrow\downarrow\right|-
       \left|\downarrow\uparrow\rangle
           \langle\downarrow\uparrow\right|\right)~,                  
\end{eqnarray}
and for the parallel pairs 
\begin{eqnarray}
-\tau^x_i&=& 
\left( \left|\uparrow\uparrow\rangle
        \langle\downarrow\downarrow\right|+
       \left|\downarrow\downarrow\rangle
        \langle\uparrow\uparrow\right|\right)~,
\nonumber\\    
-\tau^z_i&=& 
(-1)^{\sum_{j=1}^{i-1} s_i}
\left( \left|\uparrow\uparrow\rangle
       \langle\downarrow\downarrow\right|-
       \left|\downarrow\downarrow\rangle         
       \langle\uparrow\uparrow\right|\right)~,
\end{eqnarray}
the Hamiltonian (\ref{Halpha}) reduces to
\begin{eqnarray}
H_{\vec s}(\alpha)&=& -J\sum_{i=1}^{N'-1} 
\left[\alpha\tau^x_i+\tau^z_i\tau^z_{i+1} \right]      \nonumber \\
&-&J \left[\alpha\tau^x_{N'}+
     (-1)^s~\tau^z_{N'}\tau^z_{1} \right]+C_s(\alpha)~.
\label{Heffs}
\end{eqnarray}
For even $s$ the Hamiltonian (\ref{Heffs}) is simply 
the ferromagnetic QIM, but when $s$ is odd, the interaction
on the last $\{N',1\}$ bond is antiferromagnetic.

\subsection{Exact solution}

Each effective model (\ref{Heffs}) can be solved using the Jordan-Wigner 
transformation for spin operators,
\begin{eqnarray}
\tau^x_i&=&1-2 c^\dagger_i c^{}_i~,                \\
\tau^z_i&=&-\left( c^{}_i+c_i^\dagger\right)
            \prod_{j<i}\left(1-2 c_j^\dagger c^{}_j\right)~.
\label{JW}
\end{eqnarray}
Here $c_i$ is a fermionic annihilation operator at site $i$. 
After this transformation the Hamiltonian (\ref{Heffs}) becomes
\begin{eqnarray}
H_{\vec s}(\alpha)&=&
J\sum_{i=1}^{N'-1}
\left( 
    \alpha c_i^\dagger c_i 
  - c_i^\dagger  c^{}_{i+1} 
  - c^{}_{i+1} c^{}_i
  + {\rm h.c.}
\right)      \nonumber \\
&+&
J
\left( 
    \alpha c_{N'}^\dagger c^{}_{N'}
    - c_{N'}^\dagger  \tilde{c}^{}_{1} 
    - \tilde{c}^{}_{1} c^{}_{N'}
    + {\rm h.c.}
\right)      
\nonumber \\
&+&
C_s(\alpha)~,             
\label{HJW}
\end{eqnarray}
with
\begin{equation}
\tilde{c}_{1}~=~c_1~(-1)^{1+s+\sum_{j=1}^{N'}c_j^\dag c_j}~,
\label{tildec1}
\end{equation}
depending on parity of the number of $c$-quasiparticles. This parity
is a good quantum number because $c$-quasiparticles can only be
created or annihilated in pairs, as can be seen in Eq. (\ref{HJW}).   

In order to extend the sum in the first line of Eq. (\ref{HJW})to 
$i=N'$ and, at the same time, to include the second line into this 
extended sum, we split the Hamiltonian (\ref{HJW}) as
\begin{equation}
H_{\vec s}(\alpha)=P^+\,H^+_{\vec s}(\alpha)\,P^+~
                 +~P^-\,H^-_{\vec s}(\alpha)\,P^-~,
\label{Hc}
\end{equation}
where
\begin{equation}
P^{\pm}=
\frac12\left[1\pm\prod_{i=1}^{N'}\tau^x_i\right]=
\frac12\left[1\pm\prod_{i=1}^{N'}\left(1-2c_i^\dagger c_i\right)\right]
\label{Ppm}
\end{equation}
are projectors on the subspaces with even ($+$) and odd ($-$) numbers of 
$c$-fermions and  
\begin{eqnarray}
H^{\pm}_{\vec s}(\alpha)&=&
J\sum_{i=1}^{N'}\left( \alpha
  c_i^\dagger c_i^{} - c_i^\dagger  c_{i+1}^{} - 
  c_{i+1}^{}  c_i^{} 
+ {\rm h.c.}\right)      \nonumber \\
&+&C_s(\alpha)             
\label{Hpm}
\end{eqnarray}
are corresponding reduced quadratic Hamiltonians. Definition of the 
Hamiltonians (\ref{Hpm}) is not complete without boundary conditions 
which depend both on the choice of subspace $\pm$ and on the parity of 
$s$, see Eq. (\ref{tildec1}). When both $s$ and the number of 
$c$-fermions have the same parity, then the boundary condition is 
antiperiodic, i.e., $c_{N'+1}\equiv -c_1$, but when on the contrary the 
two numbers have opposite parity --- it is periodic, i.e., 
$c_{N'+1}\equiv c_1$. With these boundary conditions the Hamiltonians 
(\ref{Hpm}) and (\ref{HJW}) are the same.

The Hamiltonian (\ref{Hpm}) is simplified by a Fourier transformation,
$c_j\equiv\frac{1}{\sqrt{N'}}\sum_k c_ke^{ikj}$.
Here $k$'s are quantized pseudomomenta. In the following we assume for 
convenience that $N'$ is even (i.e., $N=4m$, $m$ integer). For periodic 
boundary conditions $k$'s take ``integer'' values (recall that $N=2N'$)
$k=0,\pm \frac{2\pi}{N'},\pm 2\frac{2\pi}{N'},\cdots,\pi~$,
and in the antiperiodic case they are ``half-integer'', i.e.,
$k=\pm\frac12\frac{2\pi}{N'},
   \pm\frac32\frac{2\pi}{N'},\cdots,
   \pm\frac12\left(N'-1\right)\frac{2\pi}{N'}~$.
As a result, the Hamiltonians (\ref{Hpm}) describe independent subspaces 
labelled by $k$, with mixed $k$ and $-k$ quasiparticle states,
\begin{eqnarray}
H^{\pm}_{\vec s}(\alpha)\!\!&=&\!\!J\sum_k
\left[2(\alpha-\cos k) c_k^\dagger c_k^{} + \sin k \left(
c_k^\dagger c^\dagger_{-k}+h.c.\right)\right]
\nonumber\\
&+&C_s(\alpha)~.
\label{Hpmk}
\end{eqnarray}

Diagonalization of $H^{\pm}_{\vec s}(\alpha)$ is completed 
by a Bogoliubov transformation,\cite{Lie61} 
$c_k~=~u_k\gamma_k+v_{-k}^*\gamma_{-k}^\dagger$, 
where the Bogoliubov modes $\{u_k,v_k\}$ are eigenmodes 
of the Bogoliubov-de Gennes equations:
\begin{eqnarray}
\epsilon u_k &=&
 2J(\alpha-\cos k)\;u_k+2J\sin k \;v_k~,         \\
\epsilon v_k &=&
 2J\sin k \;u_k-2J(\alpha-\cos k)\;v_k~. 
\end{eqnarray}
For each value of $k$ there are two eigenstates with eigenenergies 
$\epsilon_k$ and $-\epsilon_k$. Positive eigenenergies
\begin{equation}
\epsilon_k=2J\sqrt{1+\alpha^2-2\alpha\cos k}
\label{epsilonk}
\end{equation}
define quasiparticles $\gamma_k$ for each $k$ which bring the 
Hamiltonian (\ref{Hpmk}) to the diagonal form
\begin{equation}
H^{\pm}_{\vec s}(\alpha)=\sum_k \epsilon_k 
\left(\gamma_k^\dagger\gamma_k^{}-\frac12\right)+C_s(\alpha)~,
\label{Hgamma}
\end{equation}
being a sum of fermionic quasiparticles. However, thanks to the
projection operators $P^{\pm}$ in Eq. (\ref{Hc}) only states with,
even or odd numbers of Bogoliubov quasiparticles belong to the 
physical spectrum of $H_{\vec s}(\alpha)$. 

In order to find if the number of Bogoliubov quasiparticles in a given 
subspace must be even or odd we must find first the parity of the 
number of $c$-quasiparticles in the Bogoliubov vacuum in this subspace. 
This parity depends on the boundary conditions. Indeed, when they are 
antiperiodic, then $\sin k\neq 0$ for any allowed $k$ and the ground 
state of the Hamiltonian (\ref{Hpmk}) is a superposition over states 
with pairs of quasiparticles $c_k^\dag c_{-k}^\dag$. This Bogoliubov 
vacuum has even number of $c$-quasiparticles. On the other hand, for 
periodic boundary conditions we have two special cases with $\sin k=0$ 
when $k=0$ or $\pi$. When $0\leq\alpha<1$ the Bogoliubov vacuum contains 
the quasiparticle $c_0$ but not the quasiparticle $c_\pi$ and the number 
of $c$-quasiparticles is odd. In short, (anti-)periodic boundary
conditions imply (even)odd parity of the Bogoliubov vacuum.

Taking further into account that any $\gamma_k^\dag$ changes parity of 
the number of $c$-quasiparticles, we soon arrive at the simple 
conclusion that even (odd) $s$ implies that only states with even (odd) 
numbers of Bogoliubov quasiparticles belong to the physical spectrum of 
$H_{\vec s}(\alpha)$. 
Note that for odd $s$ the lowest energy state is not the Bogoliubov 
vacuum but a state with one Bogoliubov quasiparticle of minimal energy 
$\epsilon_k$. Its pseudomomentum is $k=0$ in a $+$ subspace, but in 
a $-$ subspace there is a choice between two degenerate quasiparticle 
states with $k=\pm\frac12\frac{4\pi}{N}$.

\section{Quantum phase transition}

\subsection{Correlation functions}

\begin{figure}[t!]
\includegraphics[width=8.2cm]{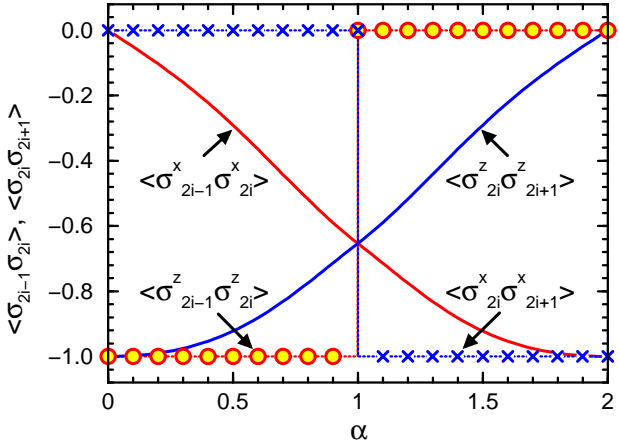}
\caption{(Color online)
Intersite pseudospin correlations on odd $\{2i-1,2i\}$ and even 
$\{2i,2i+1\}$ bonds in the XX--ZZ model 
(\ref{Halpha})--(\ref{Halpha12}) for increasing $\alpha$. 
A crossover between two types of pseudo-order, with 
$\langle\sigma_{2i-1}^z\sigma_{2i}^z\rangle=-1$ for $\alpha<1$ and
$\langle\sigma_{2i}^x\sigma_{2i+1}^x\rangle=-1$ for $\alpha>1$, 
occurs at the quantum critical point $\alpha=1$, where only 
$\langle\sigma_{2i-1}^x\sigma_{2i}^x\rangle=
 \langle\sigma_{2i}^z\sigma_{2i+1}^z\rangle=-\frac{2}{\pi}$ 
are finite. 
} 
\label{fig:jump}
\end{figure}

For any $\alpha\in(0,1)$ the ground state is found in the ${\vec s}=0$ 
(+) subspace. Pseudospin correlators in the ground state can be 
expressed by the spin correlators of the effective QIM  
Eq. (\ref{Hc}). For {\it odd pairs\/} of pseudospins: 
\begin{eqnarray}
\label{Czzodd}
\langle \sigma^z_{2i-1}\sigma^z_{2i} \rangle&=&-1~, \\
\label{Cxxodd}
\langle \sigma^x_{2i-1}\sigma^x_{2i} \rangle&=&
-\langle\tau^x_i\rangle = -1+\frac{2}{N'}\sum_k~|v_k|^2~.
\end{eqnarray}
Surprisingly, in spite of decreasing interaction $\propto (1-\alpha)$
in Eq. (\ref{Halpha}), the odd 
$\langle\sigma^z_{2i-1}\sigma^z_{2i}\rangle$ correlator (\ref{Czzodd}) 
shows {\it the same\/} perfect pseudospin order as that found at 
$\alpha=0$, while the $\langle\sigma^x_{2i-1}\sigma^x_{2i}\rangle$ one 
(\ref{Cxxodd}) gradually decreases from $0$ at $\alpha=0$ to 
$-\frac{2}{\pi}$ when $\alpha\to 1^-$ (at $N\to\infty$), 
see Fig. \ref{fig:jump}.

\begin{figure}[t!]
\includegraphics[width=8cm]{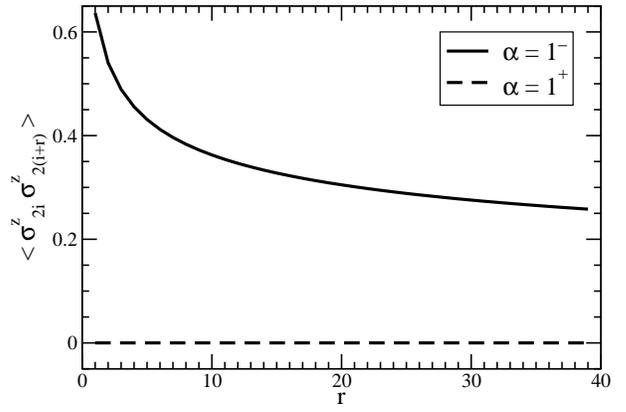}
\caption{
Distance dependence of $\langle\sigma^z_{2i}\sigma^z_{2(i+r)}\rangle$
correlator between even pseudospins belonging to different odd bonds.
Two limits $\alpha\to1^{\pm}$ demonstrate the discontinuity at the 
quantum critical point ($\alpha=1$). The correlators for $\alpha\to1^-$ 
approach the asymptotic value for large $r$ in algebraic rather than 
exponential way indicating divergent correlation length.  
} 
\label{fig:asymptotic}
\end{figure}

For {\it different\/} odd pseudospin pairs one finds ($\alpha<1$):
\begin{eqnarray}
\label{Cxxdifferent}
\langle \sigma^x_{2i-m}\sigma^x_{2j-n}\rangle &=&0~,       \\
\label{zz}
\langle \sigma^z_{2i-m}\sigma^z_{2j-n} \rangle&=&
(-1)^{m+n}\langle \tau^z_i \tau^z_j\rangle~,
\end{eqnarray}
when $i\neq j$ and for $m,n=0,1$. As is well known in the
QIM, the correlator on the right hand side of Eq. (\ref{zz}) 
is Toeplitz determinant,\cite{Bar71}
\begin{equation}
\langle \tau^z_i \tau^z_{i+r}~\rangle~=~
\left|
\begin{array}{cccc}
f_1 & f_2 & ... & f_r     \\
f_0 & f_1 & ... & f_{r-1} \\
... & ... & ... & ...          
\end{array}
\right|,
\end{equation}
with constant diagonals 
$f_r(\alpha)=\delta_{r,0}-2a_r(\alpha)+2b_r(\alpha)$, given by
$a_r(\alpha)=\frac{1}{N'}\sum_k|v_k|^2  \cos(kr)$ and
$b_r(\alpha)=\frac{1}{N'}\sum_k u_kv_k^*\sin(kr)$.
This correlator is positive for all $r$ and finite when $r\to\infty$, 
indicating long range ferromagnetic order of $\tau^z_i$ moments. When 
$r\to\infty$ it decays exponentially towards finite long range limit 
with a correlation length $\xi$ which diverges as 
$\xi\sim(1-\alpha)^{-1}$. 
As we have seen, in this limit the gap in the quasiparticle energy 
spectrum (\ref{epsilonk}) tends to $0$ and a quantum critical point
is approached in the universality class of the QIM.

For nearest neighbor {\it even pairs\/} [when $j=i+1$, $m=0$ and $n=1$ 
in Eq. (\ref{zz})], the $\langle\sigma^z_{2i}\sigma^z_{2i+1}\rangle$ 
pseudospin correlator is negative, see Eq. (\ref{zz}), as expected for 
an antiferromagnet. 
This correlator shows a complementary behavior to that of the 
$\langle\sigma^x_{2i-1}\sigma^x_{2i}\rangle$ correlator for odd bonds 
(\ref{Cxxodd}) --- it interpolates between $-1$ at $\alpha=0$ and 
$-\frac{2}{\pi}$ when $\alpha\to1^-$ (see Fig. \ref{fig:jump}). 
In contrast, the 
$\langle\sigma^z_{2i-1}\sigma^z_{2i}\rangle$ and 
$\langle\sigma^x_{2i}\sigma^x_{2i+1}\rangle$ correlators change 
in a discontinuous way at the quantum critical point ($\alpha=1$), 
where the pseudospins become entirely disordered, and
$\langle\sigma^z_{2i-1}\sigma^z_{2i}\rangle=
 \langle\sigma^x_{2i}\sigma^x_{2i+1}\rangle=0$. 
Therefore, only $\langle\sigma^x_{2i-1}\sigma^x_{2i}\rangle=
\langle\sigma^z_{2i}\sigma^z_{2i+1}\rangle=-\frac{2}{\pi}$ are finite
and contribute to the ground state energy of the compass model.

Using the symmetry of the model (\ref{Halpha})--(\ref{Halpha12}), the 
ground state correlators for $\alpha\in(1,2]$ can be obtained by mapping 
the correlators for $\alpha\in[0,1)$. In this way we obtain correlators 
that are well defined for any value of $\alpha$ except $\alpha=1$, 
where most of the correlators are {\it discontinuous\/} 
(Fig. \ref{fig:asymptotic}). For example, when $\alpha<1$ we have 
antiparallel odd pairs of pseudospins 
($\langle\sigma^z_{2i-1}\sigma^z_{2i}\rangle=-1$), but the same 
correlator tends to $0$ when $\alpha\to1^+$ (Fig. \ref{fig:jump}). 
This discontinuity of the correlators at the quantum phase transition 
is a manifestation of level crossing between ground states in different 
subspaces $\vec s$ when $\alpha\to 1^{\pm}$. 

\subsection{Ground state energy and excitations}

\begin{figure}[t!]
\includegraphics[width=8.2cm]{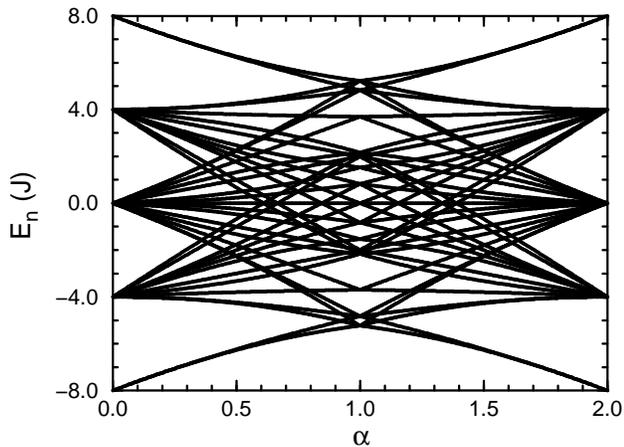}
\caption{
Eigenenergies $E_n$ of the XX--ZZ model (\ref{Halpha})--(\ref{Halpha12})
as obtained for a chain of $N=8$ sites with periodic boundary condition 
for increasing $\alpha$. Level crossing at $\alpha=1$ marks the quantum 
critical point of the compass model (\ref{H1}).
} 
\label{fig:ene}
\end{figure}

The spectrum of the Hamiltonian (\ref{Halpha}) changes qualitatively 
from a {\it ladder spectrum\/} with the width of $2JN$ at $\alpha=0$ 
to a {\it quasicontinuous spectrum\/} with the width reduced to 
$\frac{4}{\pi}JN$ at the $\alpha=1$ point (Fig. \ref{fig:ene}). 
While the ground state has degeneracy $d=2$ at $\alpha=0$ (the $+$ 
and $-$ lowest energy states in the subspace with $s=0$) and $d=1$ 
for $0<\alpha<1$ (the $+$ ground state with $s=0$), 
large degeneracy occurs at $\alpha=1$.
 
Lowest energies in different subspaces ${\vec s}$ are:
\begin{eqnarray}
{\cal E}^{\pm}_{\vec s}(\alpha)&=&
-\frac{1}{2} \,\sum_k \epsilon_k+C_s(\alpha)+{\cal O}(1)
\nonumber\\
&=& {\cal E}_0^+(\alpha)+{\cal O}(1)~,
\end{eqnarray}
where for $\alpha\le 1$
\begin{eqnarray}
{\cal E}_0^+(\alpha)&=&-\frac{1}{2}\left(1-\alpha\right)JN  \nonumber \\
&-&JN\frac{1}{2\pi}\int_0^{\pi}dk\, \sqrt{1+\alpha^2-2\alpha\cos k}
\label{E0}
\end{eqnarray}
is the energy of the ground state found in the $s=0$ ($+$) subspace
in the limit of $N\to\infty$. A similar formula to Eq. (\ref{E0}) is 
easily obtained for $\alpha\ge 1$ by a substitution 
$\alpha\rightarrow(2-\alpha)$. For any $\vec s$ the gap between 
the $+$ and $-$ lowest energy states is ${\cal O}(1/N)$, but the gaps 
$2(1-\alpha)s$ between $s=0$ and $s>0$ are ${\cal O}(1)$ and survive 
when $N\to\infty$. However, when $\alpha\to1^-$, then all ground state 
energies for $+/-$ and for any ${\vec s}$ become degenerate,
level crossing between {\it all\/} $2\times 2^{N/2}$ lowest energy 
states occurs when $N\to\infty$. This ground state degeneracy is much 
higher than $2\times 2^{L}$ found in the $L\times L$ 2D compass model,
\cite{Mil05} but the overall behavior is similar --- 
it indicates that frustrated interactions, the discontinuity in the 
correlation functions, and the {\it pseudospin liquid\/} disordered 
ground state are common features of the 1D and 2D compass models. 

\begin{figure}[t!]
\includegraphics[width=8.2cm]{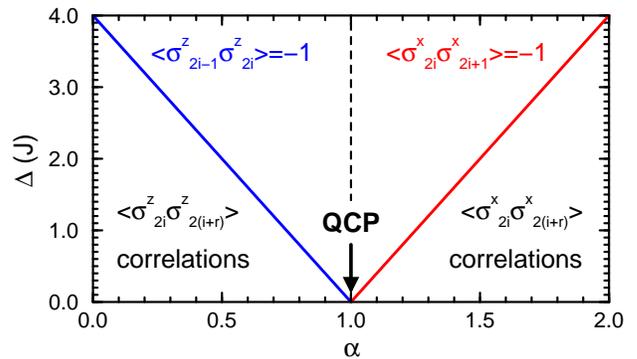}
\caption{(Color online)
Pseudospin excitation gap $\Delta$ in the XX--ZZ model 
(\ref{Halpha})--(\ref{Halpha12}) for increasing $\alpha$ (solid lines).
The gap collapses at the quantum critical point (QCP) (in the compass 
model obtained at $\alpha=1$), which separates the disordered phase with 
finite pseudospin $\langle\sigma^z_{2i}\sigma^z_{2(i+r)}\rangle$ 
correlations (left) from the one with finite 
$\langle\sigma^x_{2i}\sigma^x_{2(i+r)}\rangle$ correlations (right).
} 
\label{fig:gap}
\end{figure}

The ground state energy ${\cal E}_0^+(\alpha)$ (\ref{E0}) increases 
with $\alpha$ for $\alpha\in(0,1)$ (Fig. \ref{fig:ene}), as pseudospin 
interactions are gradually more frustrated when $\alpha\to 1$.
\cite{notexy} The ground 
state is separated from the lowest energy pseudospin excitation by 
a gap $\Delta=2\epsilon_0=4J|1-\alpha|$, which vanishes at $\alpha=1$ 
(see Fig. \ref{fig:gap}). Note that this excitation corresponds to 
reversing a $\sigma^z_i$ pseudospin component for $\alpha<1$, and a
$\sigma^x_i$ pseudospin component for $\alpha>1$.
We have verified that the linear decrease of $\Delta$ when 
$\alpha\to 1$ is well reproduced by the energy spectra obtained 
by exact diagonalization of finite systems
(as obtained, for instance, from the data of Fig. \ref{fig:ene}).

\section{Conclusions}

We have shown by an exact solution of the 1D XX--ZZ model 
(\ref{Halpha})--(\ref{Halpha12}) that pseudospin disordered states are 
triggered already by an infinitesimal admixture of the interactions 
between other spin components than those used to construct classical 
Ising models at $\alpha=0$ (\ref{H0}) or $\alpha=2$ (\ref{H2}). The 
properties of the XX--ZZ model are summarized in Fig. \ref{fig:gap}, 
with two different types of pseudospin correlations, dictated by the 
'dominating' interactions. These pseudospin correlations and finite 
excitation gap may be seen as {\it precursor\/} of the antiferromagnetic 
order induced by the type of interactions in the respective classical 
Ising model, with either $\sigma_i^z$ or $\sigma_i^x$ pseudospins,
at $\alpha=0$ or $\alpha=2$, respectively. 
These opposite trends become frustrated in the 1D compass model lying 
precisely at the quantum critical point. We anticipate that a similar
quantum critical point determines the properties of the 
{\it orbital liquid\/} state in a 2D compass model.\cite{Mil05}

In conclusion, the 1D XX--ZZ model provides a beautiful example 
of a first order quantum phase transition between two different 
disordered phases, with hidden order of pairs of pseudospins on 
every second bond. When the pseudospin interactions become balanced 
at the quantum critical point, the pseudopspin (orbital liquid) 
disordered state takes over. It is characterized by: 
(i) high $2\times 2^{N/2}$ degeneracy of ground state, and 
(ii) the gapless excitation spectrum.

\begin{acknowledgments}
We thank Peter Horsch for insightful discussions.
W.B. acknowledges kind hospitality of 
Laboratoire de Physique des Solides, Universit\'e Paris Sud, Orsay,
where part of this work was done.
J.D. was supported in part by the KBN grant PBZ-MIN-008/P03/2003
and the Marie Curie ToK project COCOS (MTKD-CT-2004-517186).
A.M.O. acknowledges support by the Polish Ministry 
of Science and Education under Project No. N202 068 32.
\end{acknowledgments}



\begin{thebibliography}{99}

\bibitem{Tok00} Y. Tokura and N. Nagaosa,
                   Science \textbf{288}, 462 (2000).

\bibitem{Fei97} L. F. Feiner, A. M. Ole\'s, and J. Zaanen,
                   \prl \textbf{78}, 2799 (1997).

\bibitem{Ole06} A. M. Ole\'s, P. Horsch, L. F. Feiner, 
                   and G. Khaliullin, 
                   \prl \textbf{96}, 147205 (2006).

\bibitem{vdB04} J. van den Brink, 
                   New J. Phys. \textbf{6}, 201 (2004).

\bibitem{Kha00} G. Khaliullin and S. Maekawa,
                   \prl \textbf{85}, 3950 (2000).

\bibitem{Kho82} K. I. Kugel and D. I. Khomskii,
		   Sov. Phys. Usp. \textbf{25}, 231 (1982).

\bibitem{Nus04} Z. Nussinov, M. Biskup, L. Chayes, and J. van den Brink,
                   Europhys. Lett. \textbf{67}, 990 (2004).

\bibitem{Nus05} Z. Nussinov and E. Fradkin,
                   \prb \textbf{71}, 195120 (2005).

\bibitem{Dou05} B. Dou\c{c}ot, M. V. Feigel'man, L. B. Ioffe, 
                   and A. S. Ioselevich,
                   \prb \textbf{71}, 024505 (2005).
		   
\bibitem{Mil05} J. Dorier, F. Becca, and F. Mila,
                   \prb \textbf{72}, 024448 (2005).

\bibitem{Kho03} D. I. Khomskii and M. V. Mostovoy,
		   J. Phys. A \textbf{36}, 9197 (2003).

\bibitem{noteaf} Generic superexchange interactions in Mott insulators  
                   are antiferromagnetic,\cite{Kho82} but the
		   present analysis can be also generalized to the
		   ferromagnetic case.
		   	   
\bibitem{Lie61} E. Lieb, T. Schultz, and D. Mattis, 
                   Ann. Phys. \textbf{16}, 407 (1961);
                S. Katsura, 
		   Phys. Rev. \textbf{127}, 1508 (1962).
		   
\bibitem{Dzi05} J. Dziarmaga, 
                   \prl \textbf{95}, 245701 (2005).

\bibitem{Aic02}	M. V. Mostovoy, D. I. Khomskii, and J. Knoester,
                   \prb \textbf{65}, 064412 (2002);
                M. Aichhorn, P. Horsch, W. von der Linden, 
		   and M. Cuoco, 
                   {\it ibid.\/} \textbf{65}, 201101 (2002).

\bibitem{Bar71} E. Barouch and B. M. McCoy, 
		   \pra \textbf{3}, 786 (1971).

\bibitem{notexy} Note that ${\cal E}_0^+(1)$ defined by Eq. (\ref{E0}) 
                   is half of the ground state energy of the full 1D 
		   XY model, see Ref. \onlinecite{Bar71}.

		   
\end{thebibliography}
\end{document}